\documentclass[prl,superscriptaddress,twocolumn]{revtex4-1}
\usepackage{amsfonts,amsmath,amssymb,graphicx,hyperref}

\newcommand{\R}{\mathbb{R}}
\newcommand{\G}{\mathcal{G}}
\newcommand{\ea}[1]{\langle #1 \rangle}
\newcommand{\set}[1]{\{ #1 \}}

\renewcommand{\L}{\mathcal{L}}
\newcommand{\e}{\varepsilon}

\newcommand{\ton}{\xrightarrow[n\to\infty]{}}

\begin{document}

\title{Clustering implies geometry in networks
}
\author{Dmitri Krioukov}
\affiliation{Northeastern University, Departments of Physics, Mathematics, and Electrical\&Computer Engineering, Boston, MA, USA}

\begin{abstract}
Network models with latent geometry have been used successfully in many applications in network science and other disciplines, yet it is usually impossible to tell if a given real network is geometric, meaning if it is a typical element in an ensemble of random geometric graphs. Here we identify structural properties of networks that guarantee that random graphs having these properties are geometric. Specifically we show that random graphs in which expected degree and clustering of every node are fixed to some constants are equivalent to random geometric graphs on the real line, if clustering is sufficiently strong. Large numbers of triangles, homogeneously distributed across all nodes as in real networks, are thus a consequence of network geometricity. The methods we use to prove this are quite general and applicable to other network ensembles, geometric or not, and to certain problems in quantum gravity.
\end{abstract}

\maketitle

In equilibrium statistical mechanics it is often possible to tell if a given system state is a typical state in a given ensemble. In network science, where statistical mechanics methods have been used successfully in a variety of applications~\cite{AlBa02,PaNe04,Gao2016Reslience}, the same question is often intractable. Stochastic network models define ensembles of random graphs with usually intractable distributions. Therefore it is usually unknown if a given real network is a typical element in the ensemble of random graphs defined by a given model, i.e., if the model is appropriate for the real data, so that it can yield reliable predictions. Progress has been made in addressing this problem in some classes of models, such as the configuration~\cite{BianconiEntropy2008,BianconiAssessing2009,BianconiEntropy2009,AnKr14,GarlaschelliLikelihood2008,GaLo09,Squartini2015Sampling} and stochastic block models~\cite{Peixoto2012Entropy,Peixoto2013Parsimonuous,Peixoto2014Hierarchical}.

Here we are interested in latent-space network models~\cite{Newman2015Generalized}. In these models, nodes are assumed to populate some latent geometric space, while the probability of connections between nodes is usually a decreasing function of their distance in this space. Latent-space models were first introduced in sociology in the 70ies~\cite{McFarland1973SocialDistance} to model homophily in social networks---the more similar two people are, the closer they are in a latent space, the more likely they are connected~\cite{McPh01}. Since then, latent-space models have been used extensively in many applications, ranging from predicting social behavior and missing or future links~\cite{Hoff2002Latent,SaCh11,Tita2014Latent}, to designing efficient information routing algorithms in the Internet~\cite{BoPa10} and identifying connections in the brain critical for its function~\cite{Gulyas2015Navigable}, to inferring community structure in networks~\cite{Newman2015Generalized}---see \cite{Barthelemy11,Bianconi2015Perspective} for surveys.

The simplest network model with a latent space is the model with the simplest latent space, which is the real line $\R^1$. Nodes are points sprinkled randomly on $\R^1$, and two nodes are connected if the distance between them on $\R^1$ is below a certain threshold~$\mu$. This random graph ensemble is known as the Gilbert model of random geometric graphs~\cite{Gilbert1961RandomPlaneNetworks,Penrose03-book}. Even in this simplest model, the ensemble distribution is intractable and unknown. Therefore it is impossible to tell if a given (real) network is ``geometric''---that is, if it is a typical element in the ensemble. One can always check (in simulations) a subset of necessary conditions: if the network is geometric, then all its structural properties must match the corresponding ensemble averages. By ``network property'' one usually means a function of the adjacency matrix. The simplest examples of such functions are the numbers of edges, triangles, or subgraphs of different sizes in the network~\cite{Orsini2015DK}. The distributions of betweenness or shortest-path lengths correspond to much less trivial functions of adjacency matrices. Since the number of such property-functions is infinite, and since their inter-dependencies are in general intractable and unknown~\cite{Orsini2015DK}, it is impossible to check if all properties match and all conditions necessary for network geometricity are satisfied. Do any sufficient conditions exist? That is, are there any structural network properties such that random networks that have these properties are typical elements in the ensemble of random geometric graphs?

Here we answer this question positively for random geometric graphs on~$\R^1$. We show that the set of sufficient-condition properties is surprisingly simple. These properties are only the expected numbers of edges~$\bar{k}$ and triangles~$\bar{t}$, or equivalently, expected degree~$\bar{k}$ and clustering~$\bar{c}=2\bar{t}/\bar{k}^2$ of every node. Specifically, we consider a maximum-entropy ensemble of random graphs in which the expected degree of every node is fixed to the same value~$\bar{k}$, while the expected number of triangles to which every node belongs is also fixed to some other value~$\bar{t}$. There is seemingly nothing geometric about this ensemble since it is defined in purely network-structural terms---edges and triangles, in combination with the maximum-entropy principle~\cite{Horvat2015Degeneracy,Squartini2015Breaking}. Yet we show that if clustering is sufficiently strong, then this ensemble is equivalent to the ensemble of random geometric graphs on $\R^1$. In general, the ensemble is not {\it sharp} but {\it soft}~\cite{Dettmann2015RGG,Penrose2013SRGG}---the probability of connections is not $0$ or $1$ depending on if the distance between nodes is larger or smaller than $\mu$, but the grand canonical Fermi-Dirac probability function in which energies of edges are distances they span on $\R^1$. Strong clustering, a fundamentally important property of real networks~\cite{Radicchi2004Communities,Radicchi2016Beyond}, thus appears as a consequence of their latent geometry.

The simplest model of networks with strong clustering is the Strauss model~\cite{Strauss1986General} of random graphs with given expected numbers of edges and triangles. The Strauss model is well studied, but many of its problematic features, including degeneracy and phase transitions with hysteresis caused by statistical dependency of edges and non-convexity of the constraints, are not observed in real networks~\cite{Horvat2015Degeneracy,Foster2010Hysteresis,Park2005Strauss}. In particular, in the Strauss model all the triangles coalesce into a maximal clique, so that a portion of nodes have a large degree and clustering close to~$1$, while the rest of the nodes have a low degree and zero clustering~\cite{Park2005Strauss,Radin2014Asymptotics}. This clustering organization differs drastically from the one in real networks, where triangles are homogeneously distributed across all nodes, modulo Poisson fluctuations and structural constraints~\cite{ColomerClustering2013,Zlatic2012Multiplicities}. If we want to fix the expected number of edges and triangles of every node to the same values $\bar{k}$ and $\bar{t}$, then the Strauss model cannot be ``fixed'' to accomplish this. Therefore instead we begin with the canonical ensemble of random graphs in which every edge $\set{i,j}$ occurs, independently from other edges, with given probability $p_{ij}$, which in general is different for different edges. This ensemble is well-behaved and void of any Strauss-like pathologies~\cite{PaNe04}. The expected degree~$\ea{k_i}$ and number of triangles~$\ea{t_i}$ at node~$i$ in the ensemble are simply~$\ea{k_i}=\sum_j p_{ij}$ and $\ea{t_i}=(1/2)\sum_{j,k}p_{ij}p_{jk}p_{ki}$. Any connection probability matrix~$\set{p_{ij}}$ satisfying constraints $\ea{k_i}=\bar{k}$ and $\ea{t_i}=\bar{t}$ for some~$\bar{k},\bar{t}$ will yield a canonical ensemble in which all nodes will have the same expected degree~$\bar{k}$ and number of triangles~$\bar{t}$. However we cannot claim that such an ensemble will be an unbiased ensemble with these constraints, because a particular matrix~$\set{p_{ij}}$ satisfying them may enforce additional constraints on the expected values of some other network properties. In other words, we first have to find a way to sample matrices~$\set{p_{ij}}$ from some maximum-entropy distribution subject only to the desired constraints.

This seemingly intractable problem finds a solution using the theory of graph limits known as graphons~\cite{LovaszBook2012}, with basic formalism introduced in network models with latent variables~\cite{CaCaDeMu02,BoPa03}. Graphon~$p(x,y)$ is a symmetric integrable function~$p:[0,1]^2\to[0,1]$, which is essentially the thermodynamic $n\to\infty$ limit of matrix~$\set{p_{ij}}$. For a fixed graph size~$n$, graphon~$p$ defines graph ensemble~$\G_n(p)$ by sprinkling $n$ nodes uniformly at random on interval~$[0,1]$, and then connecting nodes~$i$ and~$j$ with probability~$p_{ij}=p(x_i,x_j)$, where $x_{i},x_{j}$ are sprinkled positions of $i,j$ on $[0,1]$. In the $n\to\infty$ limit, the discrete node index~$i$ becomes continuous~$x\in[0,1]$. Graphs in ensemble~$\G_n(p)$ are dense, because the expected degree of a node at~$x\in[0,1]$ is~$\ea{k(x)}=n\int_0^1p(x,y)\,dy$. Here we are interested in sparse ensembles, since most real networks are sparse. Their average degrees are either constant or growing at most logarithmically with the network size~$n$~\cite{BoLaMoChHw06}. To model sparse networks, one can replace $p(x,y)$ by a rescaled graphon $p_n(x,y)=p(x,y)/n$ which depends on~$n$~\cite{BoPa03,Borgs2014LpI}. The expected degrees do not then depend on~$n$, but the number of triangles vanishes as~$1/n$, $\ea{t(x)}=(1/2n)\iint_0^1 p(x,y)p(y,z)p(z,x)\,dy\,dz$, as opposed to clustering in real networks, where it does not depend on the size of growing networks either~\cite{BoLaMoChHw06}.

The solution to this impasse is a linearly growing support of graphon~$p$. That is, let $p:\R^2\to[0,1]$ be a graphon on the whole infinite plane~$\R^2$. For any finite $n$ we simply consider its restriction to a finite square of size $n \times n$, e.g., $I_n^2$, where $I_n=[-n/2,n/2]$, so that $p_n:I_n^2\to[0,1]$ and $p_n(x,y) = p(x,y)$. Graphon $p(x,y)$ is then the connection probability in the thermodynamic limit.
In this case, both the expected degree and number of triangles at any node in the thermodynamic limit can be finite and positive: $\ea{k(x)}=\int_\R p(x,y)\,dy$ and $\ea{t(x)}=(1/2)\iint_{\R^2} p(x,y)p(y,z)p(z,x)\,dy\,dz$. For a finite graph size~$n$, the graph ensemble~$\G_n(p)$ is defined by sprinkling $n$ points~$x_i$ uniformly at random on interval~$I_n$, and then connecting nodes~$i$ and~$j$ with probability~$p_{ij}=p(x_i,x_j)$. The only difference between $\G_n(p)$ and the infinite graph ensemble~$\G_\infty(p)$ in the thermodynamic limit is that in the latter case this sprinkling is a realization~$\Pi=\set{x_{i}}$ of the unit-rate Poisson point process on the whole infinite real line~$\R$.

The main utility of using graphons here is that they allow us to formalize our entropy-maximization task as a variational problem which we will now formulate. We first observe that for a fixed sprinkling $\Pi$, the connection probability matrix $\set{p_{ij}}$ is also fixed. Since with fixed $\set{p_{ij}}$, all edges are independent Bernoulli random variables albeit with different success probabilities, the entropy of a graph ensemble $S[\G_n(p|\Pi)]$ with fixed sprinkling $\Pi$ is the sum of entropies of all edges, $S[\G_n(p|\Pi)]=(1/2)\sum_{i,j}h(p_{i,j})$, where $h(p)=-p\log p-(1-p)\log(1-p)$ is the entropy of a Bernoulli random variable with the success probability~$p$. Unfixing $\Pi$ now, the distribution of entropy $S[\G_n(p|\Pi)]$ as a function of random sprinkling $\Pi$ in ensemble $\G_n(p)$ is known~\cite{Janson2010Graphons} to converge in the thermodynamic limit to the delta function centered at the graphon entropy $s[p]$ defined below:
\begin{equation}\label{eq:graphon-entropy}
S[\G_n(p|\Pi)] \to S[\G_n(p)]\to s[p]=\frac{1}{2}\iint_{\R^2} h[p(x,y)]\,dx\,dy,
\end{equation}
where $S[\G_n(p)]$ is the Gibbs entropy of ensemble $\G_n(p)$, $S[\G_n(p)]=-\sum_{G\in\G_n(p)}P(G)\log P(G)$.
Bernoulli entropy~$S[\G_n(p|\Pi)]$ is thus self-averaging, and for large~$n$, any graph sampled from~$\G_n(p)$ is a typical representative of the ensemble. The proof in~\cite{Janson2010Graphons} is for dense graphons, but we show in the appendix that $S[\G_n(p|\Pi)]$ is self-averaging in our sparse settings as well. Therefore, our sparse ensemble~$\G_n(p)$ is unbiased if it is defined by graphon~$p^*(x,y)$ that maximizes graphon entropy~$s[p]$ above, subject to the constraints that the expected numbers of edges and triangles at every node are fixed to the same values~$\bar{k},\bar{t}$,
\begin{align}
\ea{k(x)}&=\int_\R p(x,y)\,dy=\bar{k},\label{eq:k-constraint}\\
\ea{t(x)}&=\frac{1}{2}\iint_{\R^2} p(x,y)p(y,z)p(z,x)\,dy\,dz=\bar{t}.\label{eq:t-constraint}
\end{align}

To find graphon~$p^*(x,y)$ that maximizes entropy~(\ref{eq:graphon-entropy}) and satisfies constraints~(\ref{eq:k-constraint},\ref{eq:t-constraint}), we observe that constraint~(\ref{eq:k-constraint}) implies that $p^*(x,y)$ cannot be integrable since $\iint_{\R^2}p(x,y)\,dx\,dy=\bar{k}\int_\R dx$. Therefore we first have to solve the problem for finite~$n$ and then consider the thermodynamic limit. Using the method of Lagrange multipliers, we define Lagrangian~$\L=\iint_{I_n^2}dx\,dy\,\{\frac{1}{2}h[p(x,y)]+\lambda_kp(x,y)+\frac{1}{2}\lambda_t p(x,y)\int_{I_n}p(y,z)p(z,x)\,dz\}$ with Lagrange multipliers~$\lambda_k,\lambda_t$ coupled to the degree and triangle constraints. Equation~$\delta\L/\delta p=0$ leads to the following integral equation
\begin{equation}\label{eq:integral-equation}
\log\left(\frac{1}{p(x,y)}-1\right)+2\lambda_k+3\lambda_t\int_{I_n} p(x,z)p(z,y)\,dz=0,
\end{equation}
which appears intractable. However, inspired by the grand canonical formulation of edge-independent graph ensembles~\cite{Garlaschelli2013Temperature}, we next show that for sufficiently large~$n,\bar{k},\bar{t}$, its approximate solution is the following Fermi-Dirac graphon
\begin{equation}\label{eq:FD-graphon}
p^*(x,y)=
\begin{cases}
\frac{1}{1+e^{\beta(\e-\mu)}}=\frac{1}{1+e^{2\alpha(r-1/2)}}&\text{if $0\leq r \leq 1$,}\\
\frac{1}{1+e^{\beta\mu}}=\frac{1}{1+e^{\alpha}}\equiv p^*_\alpha&\text{if $r>1$,}
\end{cases}
\end{equation}
where energy~$\e=|x-y|\geq0$ of edge-fermion $(x,y)$ is the distance between nodes~$x$ and~$y$ on $\R^1$, the chemical potential~$\mu\geq0$ and inverse temperature~$\beta\geq0$ are functions of~$\bar{k}$ and~$\bar{t}$, while $\alpha=\beta\mu$ and $r=\e/2\mu$ are the rescaled inverse temperature---the logarithm of thermodynamic activity---and energy-distance.

To show this, we first notice that if $p^*(x,y)$ is a solution, then the degree constraint~(\ref{eq:k-constraint}) becomes
\begin{equation}\label{eq:k-star}
\bar{k}=\int_{I_n} p^*(x,y)\,dy = 2\mu + p^*_\alpha(n-4\mu)\approx 2\mu + p^*_\alpha n.
\end{equation}
Therefore if the average degree $\bar{k}$ is fixed and does not depend on $n$, then $p^*_\alpha\sim1/n$ and $\alpha\sim\log n$. If $p^*_\alpha$ is small, then the last integral term in~(\ref{eq:integral-equation})---the expected number of common neighbors between nodes $x$ and $y$---is negligible for $r>1$ ($|x-y|>2\mu$), and Eq.~(\ref{eq:integral-equation}) simplifies to the equation for Erd\H{o}s-R\'enyi graphs in which only the expected degree is fixed. Its solution is constant $p^*(x,y)=1/(1+e^{-2\lambda_k})$, so that $\lambda_k = -\alpha/2$, cf.~(\ref{eq:FD-graphon}).

\begin{figure}
\includegraphics[width=2in]{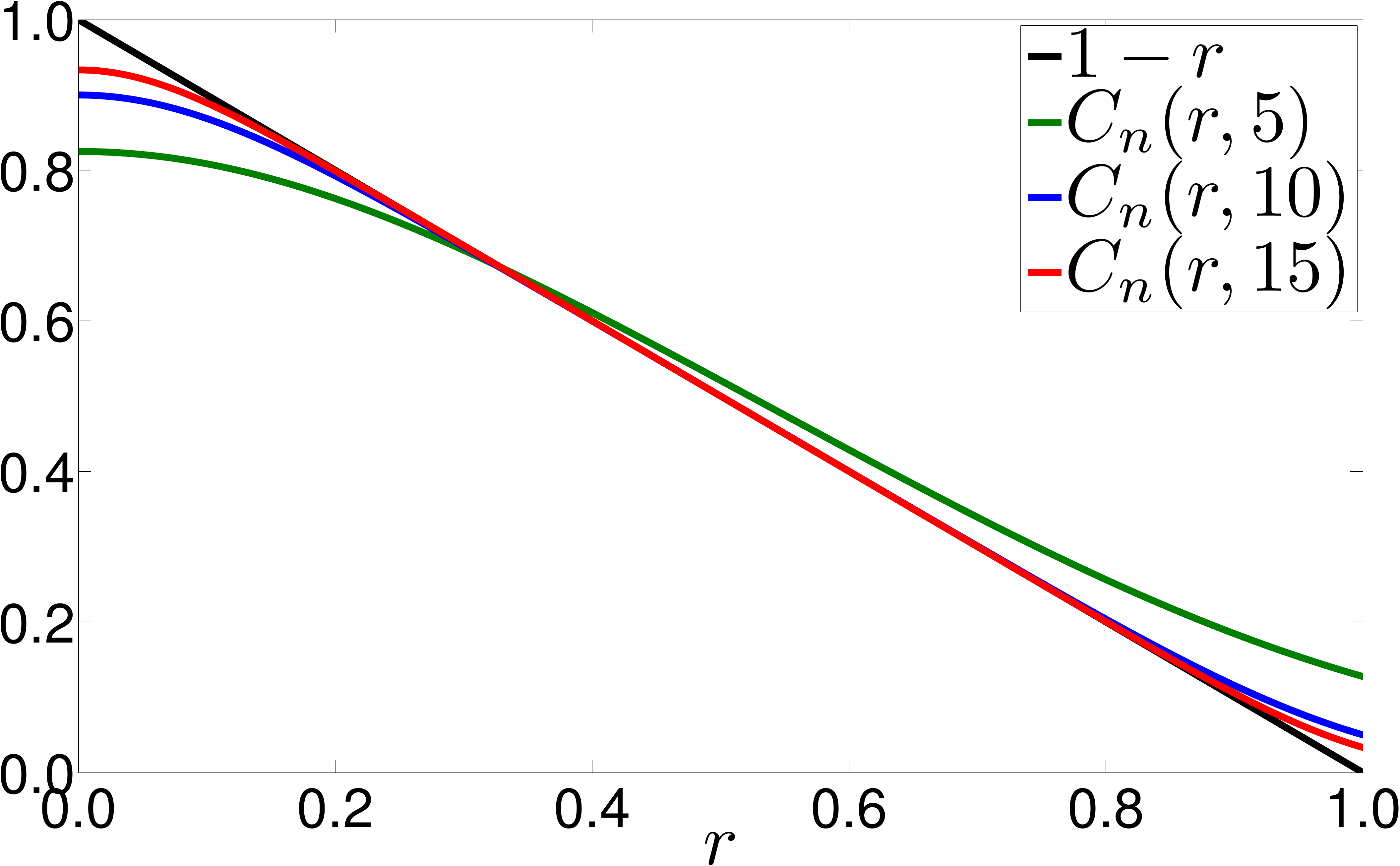}
\caption{Rescaled number of common neighbors $C_n(r,\alpha)$~(\ref{eq:common-neighbors}) versus $1-r$ for different values of~$\alpha=\beta\mu$ ($\mu=1$, $n=10^3$).
\label{fig:approx}}
\end{figure}

If $r<1$, then the common-neighbor integral in~(\ref{eq:integral-equation}) is no longer negligible, but we can evaluate it exactly for $p^*(x,y)$. The exact expression for
\begin{equation}\label{eq:common-neighbors}
C_n(r,\alpha)=\frac{1}{2\mu}\int_{I_n}p^*(x,z)p^*(z,y)\,dz,
\end{equation}
where $r=|x-y|/2\mu$, is terse and non-informative, so that we omit it for brevity. Its important property is that for large $\alpha$ it is closely approximated by $C_n(r,\alpha)\approx1-r$, Fig.~\ref{fig:approx}. In the $\alpha\to\infty$ limit this approximation becomes exact since $p^*(x,y)\to\Theta(\mu-|x-y|)=\Theta(1/2-r)$, where $\Theta()$ is the Heaviside step function---$x$ and $y$ are connected if $|x-y|<\mu$.
Approximating the common-neighbor integral in~(\ref{eq:integral-equation}) by~$2\mu(1-r)$, and noticing that $\log(1/p^*(x,y)-1)=\beta(\e-\mu)=2\alpha(r-1/2)$, we transform~(\ref{eq:integral-equation}) into
\begin{equation}
\alpha(r-1/2)+\lambda_k+3\mu\lambda_t(1-r)=0.
\end{equation}
This equation has a solution with $\lambda_k = -\alpha/2 $ and $\lambda_t=\beta/3$. This solution is consistent with the solution in the $r>1$ regime. First, the value of $\lambda_k$ is the same in both regimes $r<1$ and $r>1$. Second, one can check that the expected number of common neighbors $\int_{I_n}p^*(x,z)p^*(z,y)\,dz$ decays exponentially with $\alpha$ for any $r>1$. Therefore the common neighbor term in~(\ref{eq:integral-equation}) is indeed negligible in the $r>1$ regime, even though the prefactor $3\lambda_t=\alpha/\mu$ is large for fixed $\mu$ and large $\alpha$.

\begin{figure}
\centerline{\includegraphics{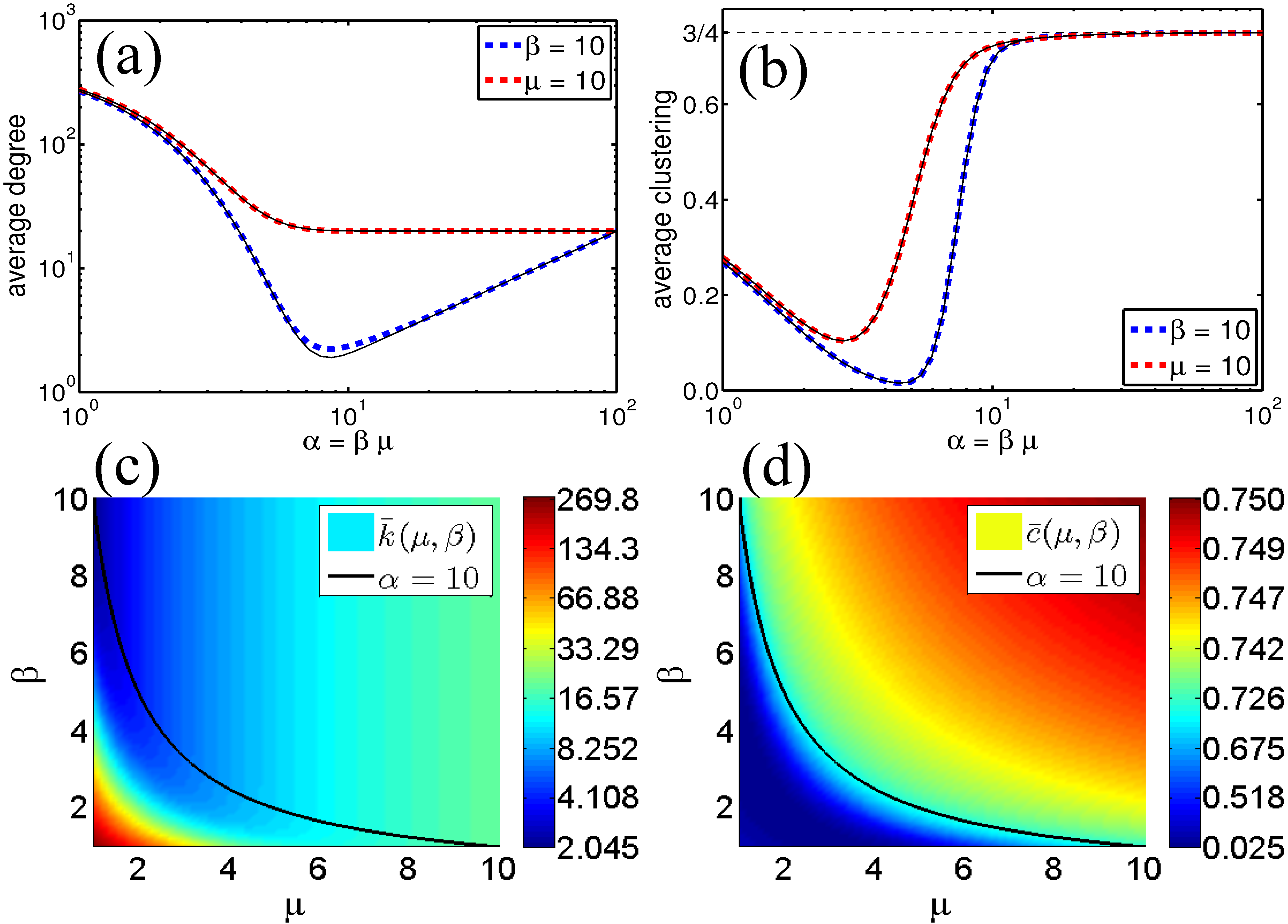}}
\caption{Average degree {\bf(a,c)} and clustering {\bf(b,d)} in soft random geometric graphs with connection probability~(\ref{eq:FD-graphon}) as functions of $\mu$ and $\beta$. The dashed curves in {\bf(a,b)} show the simulation results averaged over $100$ random graphs of size $n=10^3$ on the interval $[-500,500]$ with periodic boundary conditions. The solid curves in {\bf(a,b)} and color in~{\bf(c,d)} are the (corresponding) analytic results using~(\ref{eq:k-star}) in {\bf(a,c)}, and numeric evaluation of~(\ref{eq:c-star}) in {\bf(b,d)} with $n=10^3$. The color axes in~{\bf(c,d)} are in the logarithmic scale, with color ticks evenly spaced in $\log\bar{k}$ in~{\bf(c)} and $\log(3/4-\bar{c})$ in~{\bf(d)}.
\label{fig:kc}}
\end{figure}

Figure~\ref{fig:kc} illustrates that if $\alpha$ is large, then the expected average degree~$\bar{k}$~(\ref{eq:k-star}) and clustering
\begin{equation}\label{eq:c-star}
\bar{c}=\frac{2\bar{t}}{\bar{k}^2}=\frac{1}{\bar{k}^2}\iint_{I_n^2}p^*(x,y)p^*(y,z)p^*(z,x)\,dy\,dz
\end{equation}
in ensemble $\G_n(p^*)$ are functions of only $\mu$ and $\alpha$, respectively. Given values of the two constraints $\bar{k}$ and $\bar{t}$ (or $\bar{c}$) define the two ensemble parameters $\mu$ and $\beta$ (or $\alpha$) as the solution of Eqs.~(\ref{eq:k-star},\ref{eq:c-star}). We note that for large $\alpha$ ($\alpha>10$ in Fig.~\ref{fig:kc}), clustering is close to its maximum $\bar{c}_{\max}=3/4$ ($\bar{t}_{\max}=3\mu^2/2$), which can be computed analytically. Since our approximations are valid only for large $\alpha$, they apply only to graphs with strong clustering. In the sparse thermodynamic limit $n\to\infty$ with a finite average degree $\bar{k}$, the chemical potential $\mu$ must be finite and $\alpha$ must diverge (temperature $T=1/\beta$ must go to zero) because of~(\ref{eq:k-star}), so that only graphs with the strongest clustering are the exact solution to our entropy-maximization problem. For finite $n$ however, higher-temperature graphs with weaker clustering are an approximate solution.

We emphasize that the fact that graphon~(\ref{eq:FD-graphon}), in which the dependency on~$x$ and~$y$ is only via distance~$\e=|x-y|$, is an approximate entropy maximizer, means that the ensemble of random graphs in which the expected degree and clustering of every node are fixed to given constants, is approximately equivalent to the ensemble of soft random geometric graphs with the specific form of the connection probability, i.e., the grand canonical Fermi-Dirac distribution function that maximizes ensemble entropy constrained by fixed average energy and number of particles. In our ensemble, Fermi particles are graph edges ($0$ or $1$ edge between a pair of nodes), and their energy is the distance they span on $\R^1$. The average number of particles $\bar{m}=\bar{k}n/2$ is fixed by chemical potential $\mu$. Fixing average energy $\bar{\e}$ and fixing the average number of triangles $\bar{t}$ are equivalent because the smaller the $\bar{\e}$, the more likely the lower-energy/smaller-distance states, the larger the $\bar{t}$ thanks to the triangle inequality in $\R^1$. This equivalence explains why the Fermi-Dirac distribution~(\ref{eq:FD-graphon}) appears as an approximate solution to our entropy maximization problem constrained by fixed $\bar{k}$ and $\bar{t}$. In the zero-temperature limit $\beta\to\infty$, graphon~(\ref{eq:FD-graphon}) becomes the step function~$p^*(x,y)=\Theta(\mu-\e)$, meaning that these soft random geometric graphs become the traditional sharp random geometric graphs in which any pair of nodes is connected if their distance-energy is at most~$\mu$. All the approximations become exact in this limit.

The degree distribution in (soft) random geometric graphs is the Poisson distribution~\cite{Penrose03-book}, while in many real networks it is a power law. Triangles in real networks are still homogeneously distributed across all nodes, albeit subject to non-trivial structural constraints imposed by the power-law degree distribution~\cite{ColomerClustering2013,Zlatic2012Multiplicities,Orsini2015DK}. As shown in~\cite{SeKrBo08,KrPa10}, random geometric graphs on $\R^1$ can be generalized to satisfy an additional constraint enforcing a power-law degree distribution. This generalization still uses the grand canonical Fermi-Dirac connection probability, albeit in hyperbolic geometry, and reproduces the clustering organization in real networks. These observations lead to the conjecture that real scale-free networks are typical elements in ensembles of soft random geometric graphs with non-trivial degree distribution constraints. If so, then non-trivial community structure, another common feature of real networks, is a reflection of non-uniform node density in latent geometry~\cite{Newman2015Generalized,Zuev2015GPA}.

As a final remark we note that the graphon-based methodology we developed here is quite general and can be applied to other network models with latent variables, geometric or not, to tell if a given model is adequate for a given network. We also note that a very similar class of problems underlies approaches to quantum gravity with emerging geometry~\cite{Bianconi2015Perspective,Wu2015Emergent} where one expects continuous spacetime to emerge in the classical limit from fundamentally discrete physics at the Planck scale. Perhaps the most directly related example is the Hauptvermutung problem in causal sets~\cite{Sorkin2005Causets,BoLe87}. Given a Lorentzian spacetime, causal sets are random geometric graphs in it with edges connecting timelike-separated pairs of events sprinkled randomly onto the spacetime at the Planck density. If no continuous spacetime is given to begin with, then what discrete physics can lead to an ensemble of random graphs equivalent to the ensemble of causal sets sprinkled onto the spacetime that we observe? To answer this question, one has to solve the same ensemble equivalence problem as we solved here, except not for $\R^1$, but for the spacetime of our Universe.

\appendix

\vspace{1cm}\centerline{\large\bf Appendix}

Here we show that entropy of the considered sparse graph ensemble is self-averaging. For completeness, we first show that average entropy density converges to graphon entropy density in the thermodynamic limit, and then show that the relative variance (coefficient of variation) of the entropy distribution goes to zero in this limit. We begin with notations and definitions.

{\bf Notations and definitions.} Let $I_n=[-n/2,n/2]$ be the interval of length~$n$, and $\Pi=\set{x_i}$, $i=1,2,\ldots,n$ be $n$ real numbers sampled uniformly at random from $I_n$. For large $n$, binomial sampling $\Pi$ approximates the Poisson point process of unit rate on $I_n$. Since every $x_i$ is uniformly distributed on $I_n$, and since all $x_i$s are independent, the probability density function of sprinklings $\Pi$ is
\begin{equation}
P(\Pi)=\frac{1}{n^n}.
\end{equation}
We impose the periodic boundary conditions on $I_n$ making it a circle, so that the distance between point $i$ and $j$ is
\begin{equation}
x_{ij}=\frac{n}{2}-\left|\frac{n}{2}-|x_i-x_j|\right|.
\end{equation}
Distances $\set{x_{ij}}$ are uniformly distributed on $[0,n/2]$.

Given $\Pi$, ensemble $\G_n(p|\Pi)$ is the ensemble of graphs whose edges, or elements of adjacency matrix $\set{a_{ij}}$, are independent Bernoulli random variables: abusing notation for $p$, $a_{ij}=1$ with probability $p_{ij}=p(x_i,x_j)=p(x_{ij})$, and $a_{ij}=0$ with probability $1-p_{ij}$. There are no self-edges, so that $p_{ii}=0$. The entropy of random variable $a_{ij}$ is $h(p_{ij})$, where
\begin{equation}
h(x)=-x\log x-(1-x)\log(1-x)
\end{equation}
is the entropy of the Bernoulli random variable with success probability~$x$.
Since all $a_{ij}$s are independent in ensemble $\G_n(p|\Pi)$, its entropy is
\begin{equation}
S_n\equiv S_n(\Pi) \equiv S\left[\G_n(p|\Pi)\right]=\frac{1}{2}\sum_{i,j=1}^nh(p_{ij}),
\end{equation}
which is fixed for a given sprinkling $\Pi$. Ensemble $\G_n(p)$ is the ensemble of graphs sampled by first sampling random sprinkling $\Pi$, and then sampling a random graph from $\G_n(p|\Pi)$.

We consider entropy $S_n$ as a random variable defined by $\Pi$. This random variable is self-averaging if its relative variance vanishes in the thermodynamic limit,
\begin{equation}\label{eq:cv0}
c_v=\frac{\sqrt{\ea{S_n-\ea{S_n}}^2}}{\ea{S_n}}\ton0,
\end{equation}
where $\ea\cdot$ stands for averaging across random sprinklings~$\Pi$. We first show that average entropy density---that is, average entropy per node $\ea{S_n}/n$---converges to graphon entropy density $\sigma$,
\begin{align}
\frac{\ea{S_n}}{n}&\ton\frac{\sigma}{2},\\
\sigma&=\lim_{n\to\infty}\sigma_n=\int_\R h\left[p(x_{ij})\right]\,dx_{ij},\\
\sigma_n&=\frac{s_n}{n}=\int_{I_n}h\left[p(x_{ij})\right]\,dx_{ij},\\
s_n&=\iint_{I_n^2}h\left[p(x_i,x_j)\right]\,dx_i\,dx_j,
\end{align}
and then prove~(\ref{eq:cv0}). For notational convenience in the equations above, we have extended the support of $p(x_{ij})$ from $\R_+$ to $\R$ by $p(-x_{ij})=p(x_{ij})$.

{\bf Average ensemble entropy density converges to graphon entropy density.}
Using the definitions and observations above, we get
\begin{align}
\frac{\ea{S_n}}{n}&=\frac{1}{n}\int_{I_n^n}S_n(\Pi)P(\Pi)\,\prod_kdx_k\\
\frac{\ea{S_n}}{n}&=\frac{1}{2n}\int_{I_n^n}\sum_{i,j}h\left[p(x_i,x_j)\right]P(\Pi)\,\prod_kdx_k\label{eq:aux}\\
&=\frac{1}{2n^{n+1}}\int_{I_n^n}\sum_{i,j}h\left[p(x_i,x_j)\right]\,\prod_kdx_k.
\end{align}
The integration over $n-2$ variables $x_k$ with indices $k$ not equal to either $i$ or $j$ yields the factor of $n^{n-2}$:
\begin{equation}
\frac{\ea{S_n}}{n}=\frac{1}{2n^3}\sum_{i,j}\iint_{I_n^2}h\left[p(x_i,x_j)\right]\,dx_i\,dx_j,
\end{equation}
where we have also swapped the summation and integration.
Changing variables from $x_i$ and $x_j$ to $x_{ij}$ and $x_j$, and integrating over $x_j$ yields another factor of $n$:
\begin{equation}
\frac{\ea{S_n}}{n}=\frac{1}{2n^2}\sum_{i,j}\int_{I_n}h\left[p(x_{ij})\right]\,dx_{ij}.
\end{equation}
Since $x_{ij}$ are uniformly distributed on $[0,n/2]$, all terms in the sum contribute equally, bringing another factor of $n(n-1)\approx n^2$, the total number of terms in the sum:
\begin{equation}
\frac{\ea{S_n}}{n}=\frac{1}{2}\int_{I_n}h\left[p(x_{ij})\right]\,dx_{ij}=\frac{\sigma_n}{2}.
\end{equation}
We thus have that
\begin{equation}
\frac{\ea{S_n}}{n}\ton\frac{1}{2}\lim_{n\to\infty}\sigma_n=\frac{\sigma}{2}.
\end{equation}

{\bf Ensemble entropy is self-averaging.}
To compute
\begin{equation}
c_v=\frac{\sqrt{\ea{S_n^2}-\ea{S_n}^2}}{\ea{S_n}}
\end{equation}
we must calculate
\begin{align}
\ea{S_n^2}&=\left\langle\left(\frac{1}{2}\sum_{i,j}h\left[p(x_i,x_j)\right]\right)^2\right\rangle\\
&=I_1+I_2\text{, where}\\
I_1&=\frac{1}{4}\int_{I_n^n}\sum_{i,j}h^2\left[p(x_i,x_j)\right]P(\Pi)\,\prod_mdx_m,\\
I_2&=\frac{1}{4}\int_{I_n^n}\sum_{i,j;k,l}h\left[p(x_i,x_j)\right]h\left[p(x_k,x_l)\right]P(\Pi)\,\prod_mdx_m.
\end{align}
The first integral $I_1$ is different from~(\ref{eq:aux}) only in that instead of $h/2$ we now have $(h/2)^2$. Therefore we immediately conclude that
\begin{align}
I_1&=\frac{n}{4}\gamma_n\text{, where}\\
\gamma_n&=\int_{I_n}h^2\left[p(x)\right]\,dx.
\end{align}
If $\sigma_n=\int_{I_n}h\left[p(x)\right]\,dx$ converges to finite $\sigma$ in the $n\to\infty$ limit, as it does for the Fermi-Dirac $p^*$, then so does $\gamma_n$, $\gamma_n\ton\gamma<\infty$, because $h(p)\in[0,1]$.

To calculate the second integral $I_2$, we use $P(\Pi)=1/n^n$ and integrate over $n-4$ variables $x_m$ with indices $m$ not equal to any $i,j,k,l$, bringing the factor of $n^{n-4}$:
\begin{equation}
I_2=\frac{1}{4n^4}\sum_{i,j;k,l}\iiiint_{I_n^4}h\left[p(x_i,x_j)\right]h\left[p(x_k,x_l)\right]\,dx_i\,dx_j\,dx_k\,dx_l.
\end{equation}
Changing variables from $x_i$, $x_j$, $x_k$, and $x_l$, to $x_{ij}$, $x_j$, $x_{kl}$, and $x_l$, and integrating over $x_j$ and $x_l$, thus bringing another factor of $n^2$, we get:
\begin{equation}
I_2=\frac{1}{4n^2}\sum_{i,j;k,l}\iint_{I_n^2}h\left[p(x_{ij})\right]h\left[p(x_{kl})\right]\,dx_{ij}\,dx_{kl}.
\end{equation}
Since $x_{ij}$ and $x_{kl}$ are independent and uniformly distributed on $[0,n/2]$, every term in the double sum contributes equally, while the total number of terms is $[n(n-1)]^2\approx n^4$, yielding
\begin{align}
I_2&=\frac{n^2}{4}\iint_{I_n^2}h\left[p(x_{ij})\right]h\left[p(x_{kl})\right]\,dx_{ij}\,dx_{kl}\\
&=\left(\frac{n}{2}\int_{I_n}h\left[p(x)\right]\,dx\right)^2\\
&=\left(\frac{n}{2}\sigma_n\right)^2=\ea{S_n}^2.
\end{align}

Collecting the calculations of $\ea{S_n}$ and $\ea{S_n^2}$, we finally obtain
\begin{equation}
c_v=\frac{\sqrt{I_1+I_2-\ea{S_n}^2}}{\ea{S_n}}=\frac{\sqrt{(n/4)\gamma_n}}{(n/2)\sigma_n}=\frac{1}{\sqrt{n}}\frac{\sqrt{\gamma_n}}{\sigma_n}.
\end{equation}
If $\sigma_n\ton\sigma<\infty$, then $\gamma_n\ton\gamma<\infty$, and
\begin{equation}
c_v=\frac{1}{\sqrt{n}}\frac{\sqrt{\gamma_n}}{\sigma_n}\ton0.
\end{equation}

\vspace{1cm}

\begin{acknowledgments}
We thank G.~Lippner, P.~Topalov, M.~Piskunov, M.~Kitsak, M.~Bogu{\~n}{\'a}, S.~Horv\'at, Z.~Toroczkai, and Y.~Baryshnikov for useful discussions and suggestions. This work was supported by NSF CNS-1442999.
\end{acknowledgments}

\end{document}